# Deep optical neural network by living tumour brain cells

**Authors:** D. Pierangeli[1,4]†, V. Palmieri[2,4]†, G. Marcucci[1,4], C. Moriconi[3], G. Perini[2], M. De Spirito[2], M. Papi[2]*, C. Conti[1,4]*

**Affiliations:**

[1]Department of Physics, University Sapienza, Piazzale Aldo Moro 5, 00185 Rome (IT)

[2]Institute of Physics, Fondazione Policlinico Universitario A. Gemelli, IRCCS - Università Cattolica del Sacro Cuore, 00168 Rome (IT)

[3]School of Pharmacy and Pharmaceutical Sciences, Cardiff University, CF10 3NB Cardiff (UK)

[4]Institute for Complex Systems, National Research Council (ISC-CNR), Via dei Taurini 19, 00185 Rome (IT)

*Correspondence to claudio.conti@uniroma1.it, massimiliano.papi@unicatt.it

†These authors equally contributed

**Abstract:**

The new era of artificial intelligence demands large-scale ultrafast hardware for machine learning.[1] Optical artificial neural networks process classical and quantum information at the speed of light, and are compatible with silicon technology, but lack scalability and need expensive manufacturing of many computational layers[2–6]. New paradigms, as reservoir computing[7–9] and the extreme learning machine[10], suggest that disordered[11–16] and biological materials may realize artificial neural networks with thousands of computational nodes trained only at the input and at the readout. Here we employ biological complex systems, i.e., living three-dimensional tumour brain models, and demonstrate a random neural network (RNN) trained to detect tumour morphodynamics via image transmission[17,18]. The RNN, with the tumour spheroid[19] as a three-dimensional deep computational reservoir, performs programmed optical functions and detects cancer morphodynamics from laser-induced hyperthermia inaccessible by optical imaging. Moreover, the RNN quantifies the effect of chemotherapy inhibiting tumour growth. We realize a non-invasive smart probe for cytotoxicity assay, which is at least one order of magnitude more sensitive with respect to conventional imaging. Our random and hybrid photonic/living system is a novel artificial machine for computing and for the real-time investigation of tumour dynamics.

Brain-inspired complex systems are emerging as computing reservoirs in machine learning, optical neuromorphic computing, telecommunication, and cryptography [4,7]. This approach to learning machines does not require training of each of the deep computational layers but employs large-scale random mixing with trained input and output. This strategy is optimally suited for cheap new hardware with thousands of layers and novel applications.

Optical neuromorphic computing processes information at the speed of light[2,3], but requires a careful design and fabrication of the deep layers, which strongly hampers the development of large-scale photonic learning machines. On the other hand, random optical systems have been considered for imaging and computing at the classical and quantum level[11,17]. Recently, paradigmatic complex systems, such as spin-glasses, have been experimentally demonstrated and proposed by random photonics[12–16]. Disordered assemblies of dielectric particles furnish thousands of optical computing neurons; however, the large scattering loss limits computing applications.

Here we use real-brain cells for realizing a bio-inspired optical neural network. Our living computational reservoir has thousands of cells acting as wave-mixing nodes. Not only they furnish a large-scale computing reservoir but enable specific and very significant applications as detection of tumour morphodynamics. We use three-dimensional (3D) tumour models (3DTMs), which are largely used in oncology and are a promising platform for studying complex cell-to-cell interactions and anti-cancer therapeutics uptake and diffusion in tumour bulk. As we demonstrate here, 3DTMs disclose applications for hybrid bio/photonic computational devices.

A tumour is a population of abnormal cells with temporally unrestricted growth. Most tumours are composed of different types of cells with remarkable variety with respect to metastatic ability and chemotherapy resistance [20]. The spatial features of the tumour architecture are fundamental markers of cancer growth and invasiveness behavior [21]. For light, a 3DTM is a complex assembly of disordered scattering cells, which also evolve with time.

In our hybrid bio/photonic-hardware, the 3DTM cellular layers are the diffractive deep layers of the RNN (Figure 1). By exploiting structured light propagation,[11,18,22–24], we show that the RNN is a "universal optical interpolant" able to perform programmed functions[25]. Through external stimuli on the tumour brain cells – either of thermal or chemical nature – we control the internal weights of the living RNN and its functionality. The RNN follows subcellular cancer morphodynamics, not detected by more invasive and destructive optical imaging.

In optical neural networks, the many signals composing a laser beam [4,26] are mixed by using waveguides[2], or in free-space[3], and processed by a nonlinear operation at detection. Figure 1 compares the conventional artificial optical neural network architectures with our RNN. In the reported optical learning machines[2,3], (Figs. 1a,b), the trained internal matrix W and the biases B are implemented by tunable waveguide devices [2] or customized diffractive layers [3]. In our design, the input layer is realized by a spatial light modulator (SLM) that is iteratively trained, and the light propagates through the spherical layers of the 3DTM (Fig. 1c). The number, position, size, and optical properties (as absorption and refraction) of the complex assembly of tumour cells determine the time-dependent internal weights of the reservoir (encoded in their random transfer matrix W and variable bias B in Fig. 1b).

We use glioblastoma cells to form brain cancer models fabricated following a commonly adopted protocol[19] for 3D spheroids [Figure 1c, see Methods]. Various 3D cell culture systems have been reported, we use non-scaffolded 3D spheroid cultures[27]. If compared with in vivo tumours, the spheroid mimics heterogeneity, internal structures growth kinetic and – remarkably – drug response [21]. We consider both static and time-evolving 3DTM. In the latter case, the evolution is either induced all-optically by nonlinear optical thermal effects due to an additional infrared laser pump (used below to simulate hyperthermia) or freely evolving, due to the spontaneous cancer growth (eventually hampered by a chemotherapy, as detailed in the following).

Figure 2a shows the experimental implementation of the bio-photonic machine (see Methods). We first consider fixed tumour cells, with no spontaneous dynamics, obtained by chemical fixation. The weights of the computational network layers are static and do not change in time. By a proper training (Fig. 2b) of the input SLM, we obtain many different output images: the RNN acts as a universal optical synthesizer (Fig. 2c). In this case, once the target function is implemented, no feedback is present on the SLM weights.

The training procedure fixes the weights of the input layers but is strongly dependent on the internal structures of the deep cellular layers that are permanently fixed. If the structure of the spheroid varies, the computation gives a different result. This property allows us to track light-stimulated cancer morphodynamics with extreme sensitivity.

We first demonstrate morphodynamics sensing by inducing hyperthermia. Infrared (IR) laser-induced temperature variations in brain tissue are studied in many applications, like imaging and cancer therapy [28,29]. To induce changes in the fixed spheroid we use an IR pump beam that locally heats the cells (Methods). By affecting the internal structure of the deep layers through thermally-induced changes in the refractive index and geometry, the IR laser modifies the output of the machine (Fig. 3a). As the cancer spheroid changes morphology, the single-point target image moves and blurs, and the intensity autocorrelation decreases. By means of a recurrent approach, as the state of the deep layers changes, we retrain the SLM to keep constant the output (Fig. 3b). The RNN with feedback adapts its internal structure to maintain its functionality. This is equivalent to invert the information flow, with the constant output playing the role of an input state and the original input weights acting as the actual output. These weights detect internal changes in the tumour spheroid (Fig. 3c) and track the effect of the hyperthermia. Fig. 3d shows the tumour rearrangement time when varying the pump power.

As the transmission image results from the complex interference process in the random medium, the machine output is extremely sensitive to any small change inside the spheroid. To show that this approach outperforms standard optical methods, we repeat the hyperthermia during imaging of spheroids with confocal microscopy. We observe the cell nuclei morphology and distribution during irradiation by IR laser pulses and we did not find any detectable morphological change (Figs. 3e,f). The output of the bio-photonic neural network contains information about the entire volumetric sample and detects internal variations that are not resolved by direct imaging.

We then realize an optical machine with living tumour cells. While the tumour spontaneously evolves, we continuously adjust the input matrix on the SLM to keep the light focused in a specific point ("feedback on" case, Fig. 4a). After an initial transient, the efficiency of the operation improves, and the focus intensity increases with time. This dynamic is an outcome of the growth of tumour spheroid as confirmed by monitoring the spheroid shape with direct optical imaging in a set of parallel experiments (Fig. 4d) [30]. Figure 4c shows the persistence time $T_P$ of transmitted speckle [22] and gives the decay time on which the field decorrelates due to the tumour morphodynamics. $T_P$ scales linearly with the efficiency of the target image formation [22]. We find that $T_P$ grows with time, corresponding to an improved performance of the neural network. The enhancement of the RNN performance is related to the increased number of internal cellular layers during the growth of the 3DTM. Figure 4d shows 3DTM images ("CTRL" panels). The area of the cell assembly increases with time as shown in Figure 4e. Notably, the percentage growth in Fig. 4e of $T_p$ is orders of magnitudes larger the corresponding increase for the area in Fig. 4c.

We also obtain evidence that the living RNN track cellular processes in the 3DTM beyond the simple unconstrained growth. We consider the cytotoxic action of chemotherapeutics (Methods) [31]. We treat spheroids with cisplatin at 80μg/ml and we observe that after an initial transient the machine is still able to follow the tumour dynamics at a reduced signal level (Figs. 4b,c). This occurs because the natural growth is counter-balanced by the apoptotic effect of the cisplatin (Fig. 4e). The evolution of the optical machine output is found in correlation with direct imaging analysis (Figs. 4c,e). However, while microscopy is not able to evidence the effect of the chemotherapy, the output of the optical neural network displays significant differences (with one order of magnitude amplification) in the initial stage with respect to freely evolving tumours. The two signals in Fig. 4c are clearly separated whereas the analysis of images in Fig. 4e does not reveal statistically relevant changes. The living RNN hence furnishes real-time novel information on the tumour dynamics and is more sensitive with respect to static optical imaging.

Nowadays various methods are available to characterize tumour spheroids, including, for example, immunohistochemistry, confocal laser scanning microscopy, and two-photon microscopy [21]. All these techniques have various advantages, but also many drawbacks. The biggest limitations of the pathological analysis of tumour morphology – commonly adopted for biopsies from patients - [21] are the artifacts of the cutting process, the variability in staining and the batch effects and the lack of information about dynamics. Live imaging is being improved to follow cell dynamics, but dyes, probes, and phototoxic routines modify native cell behavior [27]. Our random optical learning machines furnish - after proper training - direct volumetric information, which is real-time, label-free, and extremely sensible, as it depends on thousands of scattering events interference processes in a 3DTM. We have demonstrated the first hybrid biological/photonic computational hardware. The use of biological matter in photonic devices is not only relevant for understanding fundamental processes in tumour dynamics but allows to explore evolving cellular assemblies as new architectures for computing. Our demonstration opens several possibilities for novel computational systems, novel diagnostic equipment, and - more in general - for the application of photonic and artificial intelligence to the study of living complex systems.

**Acknowledgments** We thank MD Deen Islam for the assistance in the laboratory. We acknowledge support from the QuantERA ERA-NET Co-fund 731373 (Project QUOMPLEX) and Sapienza Ateneo 2016 and 2017.

**Author contributions** CC and MP, conceptualization and supervision, ideation of experiments, methodology, project administration. DP, investigation by experiments and data analysis. VP, resources, and investigation by sample fabrication, imaging, and biological assays, validation. GM and MDS, methodology and theoretical analysis. CM and GP, resources by providing samples and expertise on tumour models. All the authors contributed to the writing of the original draft, review, and editing.

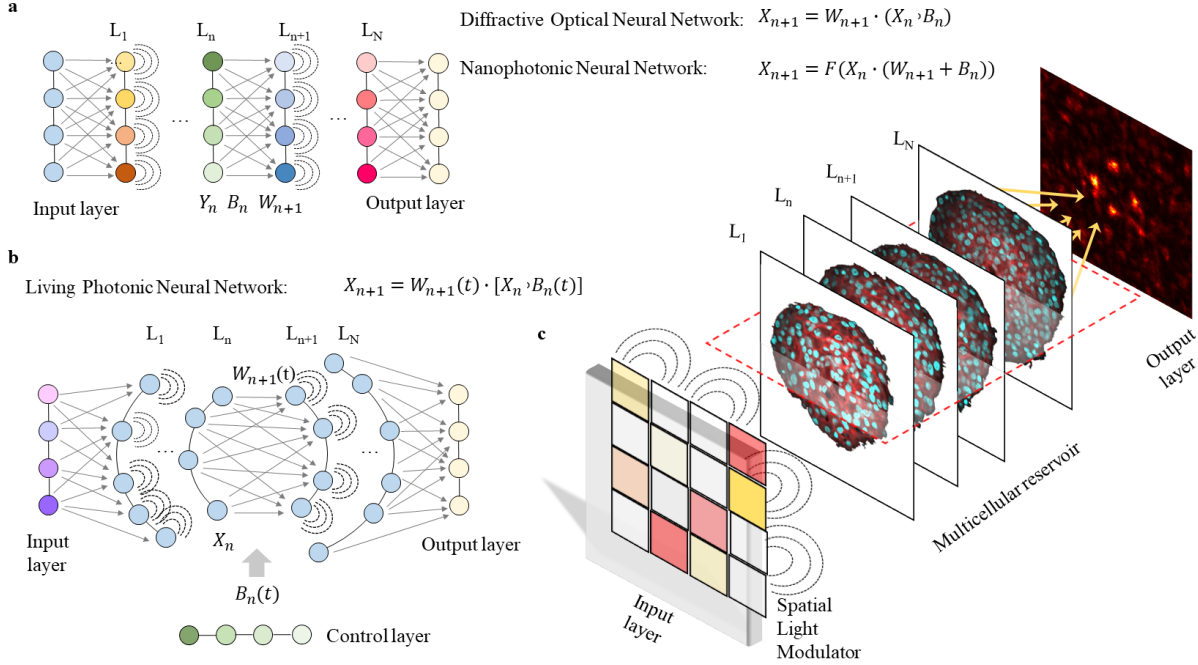

**Fig. 1. Hybrid bio/photonic deep learning machine. a**, Conventional optical deep neural networks with multiple layers ($L_1, \ldots, L_N$). Each point is a neuron with a complex transmission trained to perform a function between input and output planes. We report diffractive optical networks[3] and nanophotonic neural networks[2]. $B_n$ and $W_n$ are the bias and the linear transfer matrix at each layer n. F is a nonlinear activation function. The network is ordered, and colors encode the bias at each node. **b**, RNN with the tumour spheroid as a computing reservoir. In the input, an SLM with feedback by the output layer tailors the input signal. The internal layers are random multicellular assemblies. Each cell is a scattering center with a complex transfer function and training occurs by input SLM weights. Due to the biological nature, the internal weights change because of tumour spontaneous evolution, or under the controlled external action of an optical pump beam that heats the spheroid (non-linear control layer). This realizes a recurrent disordered network with time-dependent B and W. **c**, 3D representation of the hybrid living machine by z-stack images of the spheroid obtained after staining of cell nuclei (DAPI) and cell actin cytoskeleton (Phalloidin) and imaging by confocal microscopy.

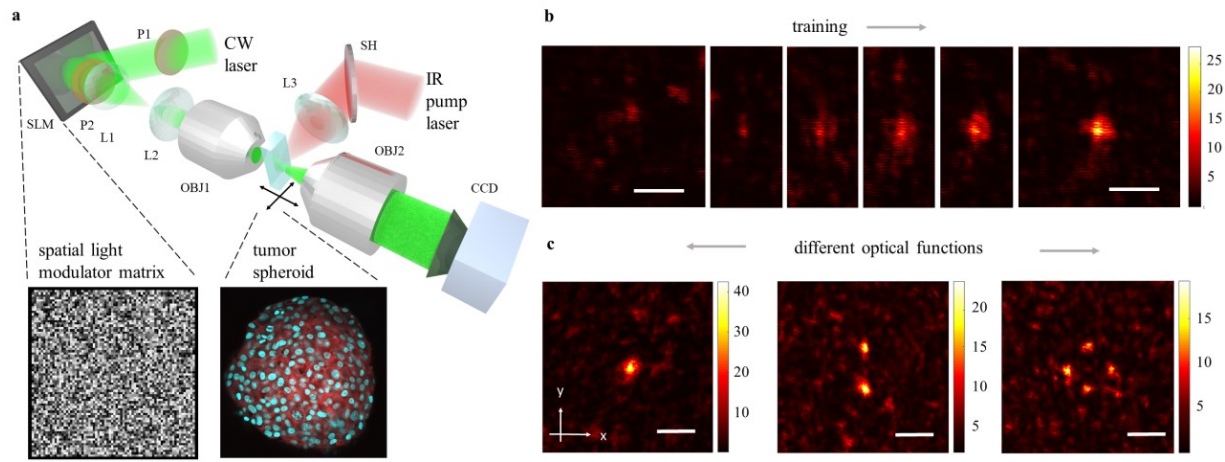

**Fig. 2. Optical synthesis with quenched tumour cells. a**, Implementation of the bio-photonic RNN. A continuous-wave (CW) laser beam (λ=532 nm) is phase-modulated by the SLM in N=72×72 addressable input modes that are mixed with different weights by the cancer spheroid. Insets show a trained phase mask on the SLM and a representative confocal image of a central plane of a glioblastoma spheroid labeled with DAPI (in cyan, marks cell nuclei) and Phalloidin (in red, marks cells actin cytoskeleton). The transmitted light distribution is detected by a CCD camera and feedback the network. **b**, Intensity distribution in the output plane during training for an input wave that is initially randomly-shaped and reaches the target shape for any random tumour sample (reservoir) by a proper fixing of the input weights. **c**, The RNN performs different optical transformations such as focusing on a single-point, two-points, and four-points targets. Intensities are normalized to the average transmission. Scale bars are 50μm. P1, P2: polarizers; L1, L2, L3: lenses; OBJ1, OBJ2: microscope objectives. SH: shutter.

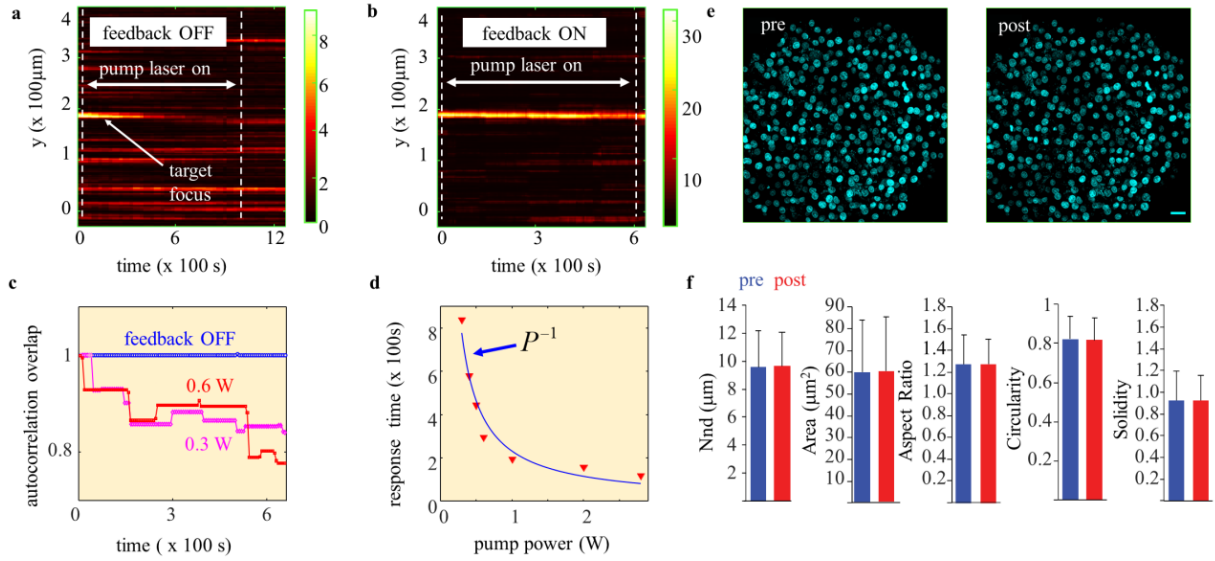

**Fig. 3. Optical neural network retrieving the tumour response to hyperthermia. a,b**, Evolution of the trained output under IR pumping: temporal dynamics of the intensity y-profile along the single-point target for (**a**) static input weights and (**b**) feedback on. Vertical dashed lines indicate the time interval of active pumping. **c**, Autocorrelation overlap of the SLM matrix in different cases: without feedback (**a**, blue dots) and with feedback (**b**) for pump power P=0.3W (magenta) and P=0.6W (red). The recurrent RNN adjusts the input weights to maintain the target operation and compensate for the spheroid structural variations. **d**, 3DTM response time varying the pump power. **e**, Confocal microscopy of spheroids labeled with DAPI before and after irradiation with IR pulses. Scale bar is 30 µm. **f**, Morphological features of cell nuclei before (blue) and after (red) irradiation with IR pulses by image analysis with FIJI software.[30] Histograms show that statistically significant alterations of nuclei nearest neighbor distances (Nnd), nuclei area, aspect ratio, circularity and solidity after hyperthermia treatment are not visible (Methods).

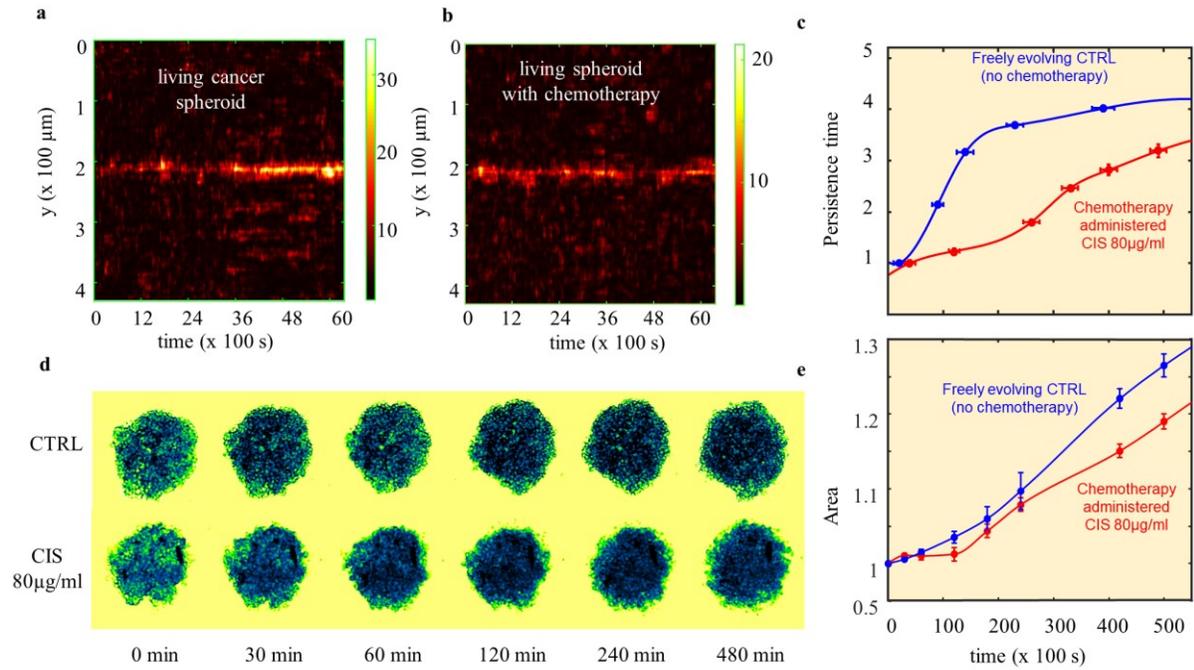

**Fig. 4. Detecting tumour growth and chemotherapy effects. a-b**, The RNN - trained to focus light in a point - operates with feedback to track (**a**) the dynamics of spontaneously growing spheroid ("control", CTRL) and (**b**) the effect of chemotherapy (80 μg/mL of cisplatin). **c**, Timescale of internal cell dynamics (persistence time $T_P$ normalized to the initial value) from the RNN output for freely evolving and chemotherapy administered spheroids. **d**, Bright-field images of untreated glioblastoma spheroids (CTRL) show a temporal increase in diameter compared to cisplatin-treated (CIS) samples. **e**, Time evolution of the area (normalized to the initial value) of the untreated and treated (80 μg/mL of cisplatin) spheroids.

## METHODS

**Tumour spheroids growth**
Spheroids have been prepared as reported in[19] using human glioblastoma cell line U-87 MG (ATCC® HTB-14™) cultured in DMEM with 10% fetal bovine serum at 37 °C, 5% $CO_2$. To form spheroids, 2500 cells have been seeded into ultra-low-adherence round-bottomed 96-well plates and immediately centrifuged (300g, 2 min) obtaining a suspended loose cell aggregate. After 4 days, tight spheroidal cell aggregates had formed. Depending on the experiments, spheroids have been analyzed in two possible states, namely fixed or living state. Fixed spheroids have been prepared with formaldehyde (3,7%) and glutaraldehyde (2,5%) mixture for 30 minutes and then washed with Phosphate Buffered Saline (PBS). Living spheroids have not been processed to monitor real-time response to therapy.

**Chemotherapy treatment**
Spheroids treated with different concentrations (10-80 μg/mL) of Cisplatin (Accord Healthcare) were imaged with a multi-well plate reader (Cytation 3, Biotek) at controlled temperature (37 °C) over time to analyze spheroid growth from brightfield images. Image analysis has been performed using INSIDIA Macro.[19]

**Optical experimental setup**
Light from a continuous-wave (CW) laser source with wavelength λ= 532nm is expanded and made to impinge on a twisted nematic liquid crystal reflective modulator (SLM, Holoeye LC-R 720). The active area of the SLM is divided into N= 72x72 independent blocks (100 pixels per block) forming the set of controlled input modes (input layer). Using a suitable combination of incident and analyzed polarizations (polarizers P1 and P2), the SLM is set into a phase-mostly modulation mode with a 10% residual intensity modulation. Phase-modulated light is imaged through a 1:3 demagnifying telescope (lens L1 and L2) on the back-focal plane of a long working distance objective (OBJ1, 50x, NA= 0.55), so that the phase of each block of the SLM matches a wavevector at the entrance of the sample. The 3mW beam is focused inside the cancer spheroid and the intensity transmitted in a plane 1.5 mm apart is imaged on a CCD camera by a collecting objective (OBJ2, 20x, NA= 0.25). During the experiments, the tumour spheroid is embedded in a physiological water solution and kept at room temperature. The morphological structure of the quenched tumour spheroid, approximately 450μm sized, has been fixed during growth, which makes the biological sample optically stationary over the measurement time (several minutes). A CW near-infrared laser source (λ = 1064nm) is used as a pump beam to induce cancer morphodynamics (laser-induced hyperthermia) by means of thermo-mechanical effects mediated by the absorption of water molecules.

**Training method**
The generic input state is a combination of the N optical modes with arbitrary phases from 0 to 2π and gives a low-intensity speckle pattern as output (Fig. 2b). Training of the optical random neural network is achieved by a feed-forward algorithm that minimizes a cost function expressed in terms of a target local intensity distribution in the transmission plane (output layer). Specifically, at each iteration of the algorithm we randomly select a cluster of input modes and adjust the corresponding phases with π/5 resolution on the SLM; the change is stored only if the cost function decreases, that is, if the intensity detected on specific groups of the CCD pixel (target) increases.
As the algorithm converges, random clusters of decreasing size are selected to avoid trapping into local computational minima. In the case of multiple-points targets, at each iteration, the input matrix change is stored only if combinations of cost functions associated to each single-point target are simultaneously minimized. After the training stage, that in Fig. 2b we limit to approximately $10^3$ iterations, the input layer encodes the proper distribution of modes (input weights) that mixed by the deep multicellular layers performs the specific operation. These photonic input-output functions are extremely sensitive to changes in the reservoir structure (the multicellular cancer network) and can be exploited for monitoring and sensing its morphology. When the optical neural network operates with feedback on, this training algorithm runs while the cancer spheroid changes morphology over time, adapting the input phases on the SLM to maintain the output target.

**Confocal microscopy imaging and analysis**
To verify if light treatment impaired spheroid structure, the hyperthermia experiment was performed on fixed spheroids stained with 4',6-diamidine-2-fenilindolo (DAPI, Sigma Aldrich) under a Confocal Microscope (Nikon A1 MP). Z-stacks of each spheroid have been acquired (Δy=1 μm) and analyzed with FIJI software.[30] A set of nuclei morphology describers have been retrieved from confocal images, i.e. nearest neighbor distance, area, circularity ($4\pi*Area/Perimeter^2$), aspect ratio of the nucleus fitted ellipse (Major axis/Minor axis) ranges from 0 (infinitely elongated polygon) to 1 (perfect circle).   The aspect ratio of the particle's fitted ellipse (Major Axis/Minor Axis), and solidity (Area/convex area). Labeling of actin cytoskeleton has been obtained by Alexa Fluor 647 phalloidin (Life Technologies), according to the manufacturer protocol.

**Data and materials availability**
The authors declare that the data supporting the findings of this study are available within the paper.

**Statistical analysis**
Each experiment has been repeated on biological replicates (n=3) with 3 technical replicates for each sample. Data statistics were analyzed by calculating the t-test between two groups, and One-way analysis of variance (ANOVA) for multiple groups with post-hoc Bonferroni test (Graphpad Prism v7, GraphPad Software Inc.). Unless otherwise noted, all results were expressed as the mean ± s.d. A value of $p < 0.05$ was considered statistically significant. Power analysis was not conducted to determine sample size and investigators were not blinded.